\documentclass[twocolumn,10pt,aps,pre,superscriptaddress]{revtex4-2}
\usepackage{hyperref}
\usepackage{graphicx, float}
\usepackage{amsfonts, amssymb, amsmath, wasysym, mathrsfs}
\usepackage{tabularx, multirow}
\usepackage{xcolor}
\usepackage{color}

\newcommand{\omegabold}{\boldsymbol\omega}

\newcommand{\corr}[1]{{\color{black} #1}}

\begin{document}

\title{Drift towards isotropization during the 3D hydrodynamic turbulence onset}

\author{D.\,S.~Agafontsev}
\affiliation{Shirshov Institute of Oceanology of RAS, Moscow, Russia}
\affiliation{Skolkovo Institute of Science and Technology, Moscow, Russia}
\affiliation{P.N. Lebedev Physical Institute of RAS, Moscow, Russia}

\author{A.\,S.~Il'yn}
\affiliation{P.N. Lebedev Physical Institute of RAS, Moscow, Russia}
\affiliation{National Research University Higher School of Economics, Moscow, Russia}

\author{A.\,V.~Kopyev}
\affiliation{P.N. Lebedev Physical Institute of RAS, Moscow, Russia}

\begin{abstract}
The incompressible three-dimensional Euler equations develop very thin pancake-like regions of exponentially increasing vorticity. 
The characteristic thickness of such regions decreases exponentially with time, while the other two dimensions do not change considerably, making the flow near each pancake strongly anisotropic. 
The pancakes emerge in increasing number with time, which may enhance the anisotropy of the flow, especially if they orient similarly in space. 
In the present paper, we study numerically the \corr{anisotropy} by analyzing the evolution of the so-called \textit{isotropy markers} [Phys. Rev. Fluids 10, L022602 (2025)]. 
We show that these functions drift slowly towards unity, indicating the process of slow isotropization, which takes place without the viscous scales getting exited and despite the similar orientation of the emerging pancakes.
\end{abstract}

\maketitle


\section{Introduction}

The research of turbulence remains essential, given its ubiquity in natural systems and its considerable resistance to analytical treatment~\cite{falkovich2001particles,dimotakis2005turbulent}. 
The developed stationary turbulence is the most studied case and serves as a starting point for many theoretical frameworks in the field. 
A key element in constructing its theory is the Kolmogorov hypothesis of the local isotropy of the velocity field~\cite{kolmogorov1941local,kolmogorov1941dissipation,biferale2005anisotropy}.  Experiments~\cite{moisy1999kolmogorov} show that a developed turbulent flow is well described within the model of stationary isotropic turbulence with additive forcing~\cite{novikov1965functionals}. 
In the theory of developed turbulence, viscosity plays a central role, allowing the turbulent process to reach a steady state~\cite{frisch1999turbulence,benzi2023lectures}.

However, the turbulence onset from a large-scale initial disturbance is a much less studied problem. 
In this case, the viscosity can be neglected and the Euler equations can be used for analysis. 
Then, the description of the initial stages of turbulence formation is an essentially non-stationary problem both in time and in space~\cite{alexakis2018cascades}, and the assumption of local isotropy, especially for anisotropic initial flows, does not appear as natural as in the case of the Kolmogorov turbulence. 

Since the early numerical experiments by Brachet \textit{et al}~\cite{brachet1992numerical}, it is known that, at the turbulence onset, the flow \corr{governed by the Euler equations} develops regions of exponentially increasing vorticity in the form of pancakes (thin vorticity sheets). 
In contrast, the stationary turbulence \corr{described by the Navier-Stokes equations} is characterized by \corr{significantly different types of structures, e.g., filaments~\cite{jimenez1998characteristics, farge2001coherent, zybin2015model}}, confirming that the statistical properties of the flow at this stage are essentially different. 
\corr{In particular, the emergence and evolution of pancakes in Eulerian dynamics, their destruction when the viscous scales get excited and emergence of filaments (ropes) during transition to a fully developed turbulent state have been recently observed numerically in~\cite{krstulovic2024initial}.

At the stage of Eulerian dynamics,} the characteristic pancake thickness decreases exponentially with time, while the other two dimensions do not change significantly~\cite{agafontsev2015development,agafontsev2016asymptotic,agafontsev2018development}. 
The number of pancakes increases, and together they provide the leading contribution to the energy spectrum. 
When the pancakes orient similarly in space, representing strongly anisotropic configuration of the flow, the gradual formation of the Kolmogorov energy spectrum $E_{k}\propto k^{- 5/3}$ and the corresponding power-law scalings for the structure functions of velocity have been observed~\cite{agafontsev2015development,agafontsev2016development,agafontsev2019statistical}, in a fully inviscid flow. 
Conversely, when the pancakes orient randomly, the power-law scalings are not observed throughout the simulation time. 
It is natural to suggest, therefore, that anisotropy promotes the formation of the Kolmogorov energy spectrum and may enhance as turbulence develops, until the viscous scales will come into play.

The aim of the present paper is to numerically test the latter hypothesis, namely, whether the anisotropy is enhancing during the turbulence onset. 
We use a new approach developed recently in~\cite{il2025stochastic} for the analysis of \corr{anisotropy} through a system of \corr{non-dimensional} stochastic identities. 
These identities \corr{involve} correlators of certain combinations composed of components of second-rank tensors. 
The correlators can be used as \textit{isotropy markers}: the closer the correlators are to unity (their value in a fully isotropic case), the closer the statistics is to isotropic. 


Specifically, we perform simulations of the 3D Euler equations taking initial conditions as a superposition of a shear flow $\mathbf{v}=(\sin z, \cos z, 0)$ and a random periodic flow; \corr{the simulation time interval is limited from above to model the 3D Euler dynamics accurately}. 
The presence of the shear flow influences the orientation of the emerging pancakes: from fully random, when this flow is absent, to almost unidirectional near the $z$-axis, when it dominates. 
We observe that, for a random initial flow, the isotropy markers remain sufficiently close to unity at all times. 
For approximately equal mixtures of the shear and random flows, the markers also remain practically unchanged over time, but some of them may have values significantly different from unity. 
When the shear flow dominates, all of the initial flows we have studied led to a nearly unidirectional pancake orientation, while the isotropy markers, initially deviating from unity by orders of magnitude, demonstrated a slow drift, similar to an exponential decay, towards unity. 

The paper is organized as follows. 
In Section~\ref{sec:markers}, following the results obtained in~\cite{il2025stochastic}, we introduce the isotropy markers. 
In Section~\ref{Sec:NumMethods} we discuss our numerical methods and in Section~\ref{Sec:Results} we demonstrate our main results. 
The concluding remarks are given in Section~\ref{Sec:Conclusions}.


\section{Isotropy markers}
\label{sec:markers}

Consider an arbitrary, non-degenerate, three-dimensional second-rank tensor $\mathbf{A}$, and the associated symmetric, positive-definite quadratic form $\mathbf{\Gamma} = \mathbf{A}^T \mathbf{A}$. 
In a given basis, the Gram matrix of the latter can be represented via the Gauss decomposition of the former,
\begin{equation}
    \mathbf{\Gamma} = \mathbf{Z}^T \mathbf{D}^2 \mathbf{Z},
    \label{E:Gamma}
\end{equation}
where $\mathbf{D} = \mathrm{diag}\{D_{xx}, D_{yy}, D_{zz}\}$, and $\mathbf{Z}$ is an upper-triangular matrix with unity on its main diagonal.
Note that the elements of the diagonal matrix $\mathbf{D}$ do not coincide generally with the eigenvalues of $\mathbf{\Gamma}$ and depend on the chosen basis. 
As follows from Eq.~(\ref{E:Gamma}), 
\begin{equation}
    D_{xx}^2 = \Gamma_{xx}, \quad D_{yy}^2 = \frac{m_{xy}}{\Gamma_{xx}}, \quad D_{zz}^2 = \frac{\det \boldsymbol{\Gamma}}{m_{xy}},
    \label{E:Jacobi}
\end{equation}
where $m_{xy}^2$ is the second leading principal minor of $\boldsymbol{\Gamma}$.

Now, let $\mathbf{A}$ be random and its probability measure be invariant under rotations (isotropic),
\begin{equation}
    \rho(\mathbf{A}) = \rho(\mathbf{O}^{-1} \mathbf{A} \mathbf{O}), \quad \forall\, \mathbf{O} \in O(d).
\end{equation}
Then, the following stochastic identities hold~\cite{sirota2024lagrangian, il2025stochastic},
\begin{eqnarray}
    \left\langle D_{zz}/D_{yy} \right\rangle &=& \left\langle \Gamma^{1/2}_{xx}\,(\det \boldsymbol{\Gamma})^{1/2}\, /\, m_{xy} \right\rangle = 1, \nonumber\\
    \left\langle D_{yy}/D_{xx} \right\rangle &=& \left\langle m_{xy}^{1/2} / \Gamma_{xx} \right\rangle = 1, \label{identities}\\
    \left\langle D_{yy}D_{zz} / D_{xx}^2 \right\rangle &=& \left\langle (\det\boldsymbol{\Gamma})^{1/2} / \Gamma_{xx}^{3/2} \right\rangle = 1, \nonumber
\end{eqnarray}
\corr{where $\langle ...\rangle$ means averaging over the volume of the flow.} 
The first of these identities requires only the axial symmetry for the probability measure around the $x$-axis, while the second identity requires the same around the $z$-axis. 
The identities~(\ref{identities}) hold for arbitrary isotropic probability measures at all times: their existence is completely determined by the geometric properties of the rotation group $O(3)$, independent of the dynamics.


As we have noted, the elements of the diagonal matrix $\mathbf{D}$ depend on the chosen basis, so that, by relabeling the axes, the identities~(\ref{identities}) can be extended,
\begin{eqnarray}
    \left\langle \Gamma_{ii}^{1/2} (\det \boldsymbol{\Gamma})^{1/2} \,/\, m_{ij} \right\rangle &=& 1, \label{m1}\\
    \label{m2}
    \langle m_{ij}^{1/2} / \Gamma_{ii} \rangle &=& 1, \label{m2}\\
    \label{m3}
    \left\langle (\det \boldsymbol{\Gamma})^{1/2} / \Gamma_{ii}^{3/2} \right\rangle &=& 1, \label{m3}
\end{eqnarray}
where $i,j = x,y,z$, $i\neq j$, and the summation over repeated indices is \textit{not} implied. 
For the identity~(\ref{m1}) to hold, only the axial symmetry around the $i$-axis is sufficient, while for the identity~(\ref{m2}) it is sufficient to have axial symmetry around the $k$-axis, $k \neq i, j$.

As has been demonstrated in~\cite{il2025stochastic}, in the case of anisotropic flows, the correlators~(\ref{m1})-(\ref{m3}) can be used as \textit{isotropy markers}. 
In particular, for a channel flow, the markers calculated for the tensor of velocity gradients $A_{ij} = \partial v_{i}/\partial x_{j}$ equal unity near the axis of the channel, where the flow is close to isotropic. 
On the contrary, near the walls of the channel, the markers exhibit systematic variation directly related to the shear flow.

In this work, we also study anisotropy using the Gram matrix constructed from the velocity gradients tensor. 
Then, the explicit expressions for the matrix minors can be written \corr{as follows,
\begin{eqnarray}
    \Gamma_{ij} &=& \frac{\partial v_\alpha}{\partial x_i} \frac{\partial v_\alpha}{\partial x_j}, \\
    m_{ij} &=& \Gamma_{ii}\Gamma_{jj} - \Gamma_{ij}^{2},
\end{eqnarray}
where} the summation over the Greek indices is implied, while no summation is performed over the Latin indices. 
\corr{Due to the relation $\mathbf{\Gamma} = \mathbf{A}^T \mathbf{A}$, it is easy to see that the diagonal elements $\Gamma_{ii}$ and the determinant $\det \boldsymbol{\Gamma}$ are positive definite; therefore, the minors $m_{ij}$ are also positive definite according to the Sylvester's criterion.}


\section{Numerical methods}
\label{Sec:NumMethods}

We perform numerical simulations of the 3D Euler equations in the vorticity formulation,
\begin{equation}
	\frac{\partial\omegabold}{\partial t} = \mathrm{rot}\,(\mathbf{v}\times \omegabold), \quad \mathbf{v} = \mathrm{rot}^{-1}\omegabold,
	\label{Euler2}
\end{equation}
in the periodic box $\mathbf{r} = (x,y,z)\in [-\pi ,\pi ]^{3}$ using the pseudo-spectral Runge-Kutta fourth-order method. 
The inverse of the curl operator and all the spatial derivatives are calculated in the Fourier space. 
We use an adaptive anisotropic rectangular grid, which is uniform for each direction and adapted independently along each of the three coordinates; the adaption comes from analysis of the Fourier spectrum of vorticity. 
Time stepping is implemented via the CFL stability criterion with the Courant number $0.75$. 
We start with the cubic grid $128^3$, refine it until the total number of nodes reaches $1024^{3}$, then fix the grid and continue until the Fourier spectrum of vorticity at $2K_{\max}^{(j)}/3$ exceeds $10^{-10}$ times its maximum value along any of the three directions. 
Here $K_{\max}^{(j)}=N_{j}/2$ is the maximum wavenumber and $N_{j}$ is the number of nodes along directions $j=x,y,z$. 
\corr{The simulations are then stopped in order to consider only the evolution that accurately reflects the 3D Euler dynamics; in terms of vorticity field in the physical space, our stopping criterion} corresponds to a resolution of the finest vorticity structure with about $8$-$10$ grid nodes at the full width at half maximum. 

For more details, we refer the reader to~\cite{agafontsev2015development, agafontsev2018development, agafontsev2016development}, where it has been verified that the accuracy within the simulation time interval is very high. 
In particular, (i) the energy and helicity are conserved up to a relative error smaller than $10^{-11}$, (ii) the results of the simulations carried out on larger grids match those obtained on smaller grids, and (iii) the simulations in the so-called vortex line representation, which is a partial integration of the Euler equations with respect to conservation of the Cauchy invariants~\cite{KuznetsovRuban}, produce the same vorticity field.

The initial conditions are taken as a sum of a shear flow and a random periodic flow. 
The shear flow, 
\begin{equation}
	\omegabold_{\mathrm{sh}}(\mathbf{r}) = (\sin z, \cos z, 0), \quad |\omegabold_{\mathrm{sh}}(\mathbf{r})| = 1,
	\label{IC-shear}
\end{equation}
represents a stationary solution of the Euler equations with velocity equal to vorticity, $\mathbf{v}_{\mathrm{sh}} = \omegabold_{\mathrm{sh}}$. 
This component is ``maximally anisotropic'': it has non-zero gradients only along the $z$-axis. 
The random flow is determined as a truncated Fourier series, 
\begin{eqnarray}
	\omegabold_{\mathrm{r}}(\mathbf{r}) = \sum_{\mathbf{h}}\bigg[\mathbf{A}_{\mathbf{h}}\cos(\mathbf{h}\cdot\mathbf{r}) + \mathbf{B}_{\mathbf{h}}\sin(\mathbf{h}\cdot\mathbf{r})\bigg],
	\label{IC-random}
\end{eqnarray}
where $\mathbf{h} = (h_x, h_y, h_z)\in[-2,2]^{3}$ is a vector with integer components. 
The real vectors $\mathbf{A}_{\mathbf{h}}$ and $\mathbf{B}_{\mathbf{h}}$ are chosen as random numbers with zero mean and variance $\sigma_{\mathbf{h}}^{2} = g(\mathbf{h})$, satisfying the orthogonality conditions $\mathbf{h}\cdot \mathbf{A}_{\mathbf{h}} = \mathbf{h}\cdot \mathbf{B}_{\mathbf{h}} = 0$ required for incompressibility. 
The function $g(\mathbf{h}) = e^{-(h-h_{0})^{2}/\mu^{2}}$, $h=|\mathbf{h}|$, with $h_{0}=\sqrt{2}$ and $\mu = 1/\sqrt{8}$ for the variance, promotes the presence of harmonics of the form $(1, 1, 0)$ and $(1, 1, 1)$ and suppresses other harmonics, e.g., $(1, 0, 0)$ and $(2, 0, 0)$, to create ``more isotropic'' flows. 
The zeroth harmonic is erased, $\mathbf{A}_{\mathbf{0}} = \mathbf{B}_{\mathbf{0}} = \mathbf{0}$, and, for simplification, we additionally erase all harmonics smaller than $0.05$ times the amplitude of the maximum harmonic. 

The resulting initial flow represents a combination,
\begin{eqnarray}
	t=0: \quad \omegabold(\mathbf{r}) = a_{0}\bigg((1-M)\,\omegabold_{\mathrm{sh}} + R \,M\,\omegabold_{\mathrm{r}}\bigg), \label{IC-mix}
\end{eqnarray}
where $M\in[0,1]$ is the mixing coefficient that regulates the level of anisotropy, $R = \sqrt{4\pi^3/E_{r}}$ is the coefficient that rescales the random component $\omegabold_{\mathrm{r}}$ to the same total energy $E_{\mathrm{sh}} = \int \frac{v_{\mathrm{sh}}^2}{2}\,dV = 4\pi^{3}$ as has the shear flow~(\ref{IC-shear}) in the box $\mathbf{r}\in[-\pi,\pi]^{3}$, $E_{r}$ is the total energy of the random flow $\omegabold_{\mathrm{r}}$ in the same box, and $a_{0}$ is the amplitude of the mix designed to make the maximum vorticity $\omega_{\max}=\max_{\mathbf{r}}|\omegabold|$ close to unity. 


\section{Results}
\label{Sec:Results}

We first investigate the evolution of isotropy markers for simulation, which starts from a generic periodic initial flow constructed with $M=1$ and $a_{0}=0.32$; hereafter we denote this flow as $\mathrm{IF_{1}}$. 
The corresponding simulation ended at the final time $t_{f} = 6.32$ on a practically isotropic grid $1024\times 972\times 972$ and demonstrated a $3.3$-fold increase in the global maximum of vorticity from its initial value $\omega_{\max}=1$ to its final value $3.3$. 
At the end of the simulation, we identified $47$ pancake structures, which are oriented in space without any visible preference. 
Note that, when working with pancakes, we repeat the steps described in detail in~\cite{agafontsev2015development,agafontsev2016asymptotic,agafontsev2016development,agafontsev2022compressible} and for this reason do not describe them here for brevity.

\begin{figure}[t]\centering
	\includegraphics[width=9cm]{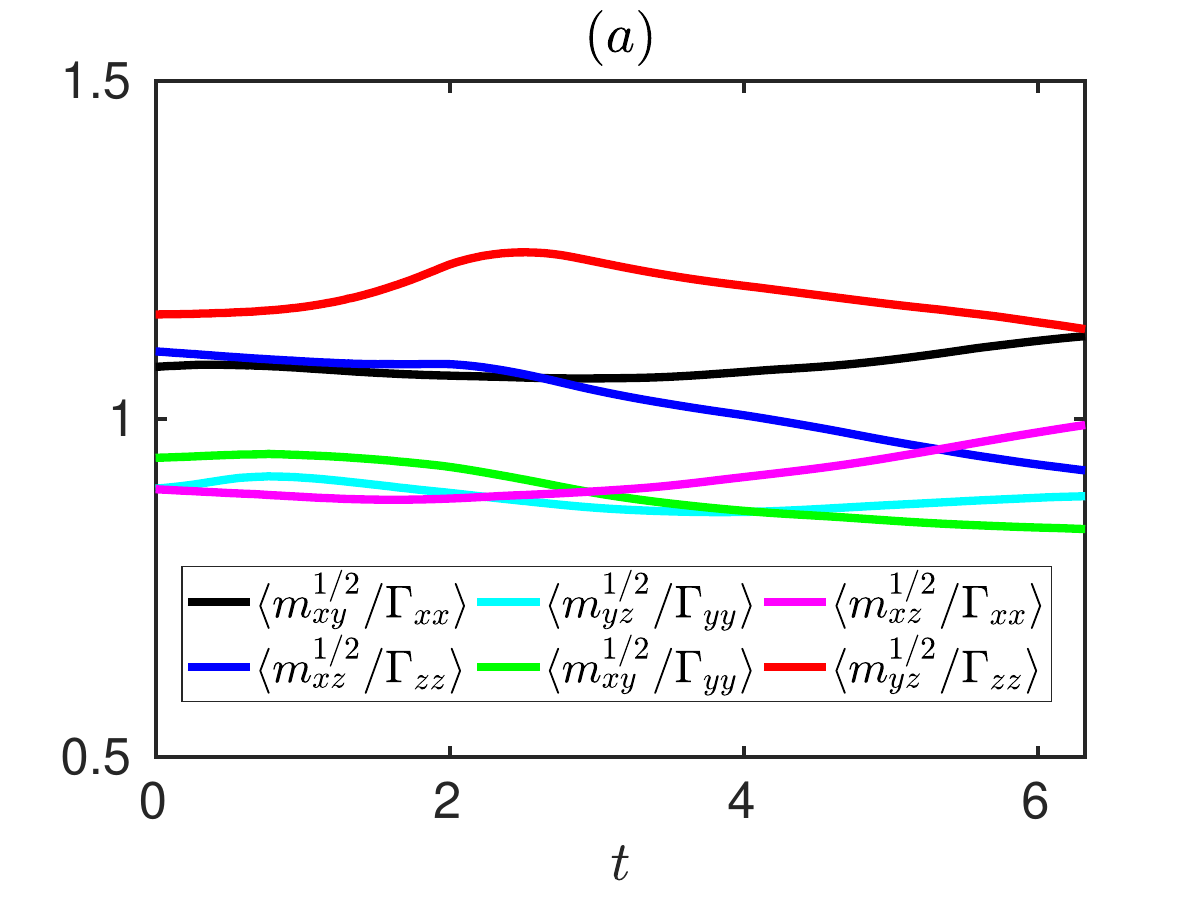}\\
	\includegraphics[width=9cm]{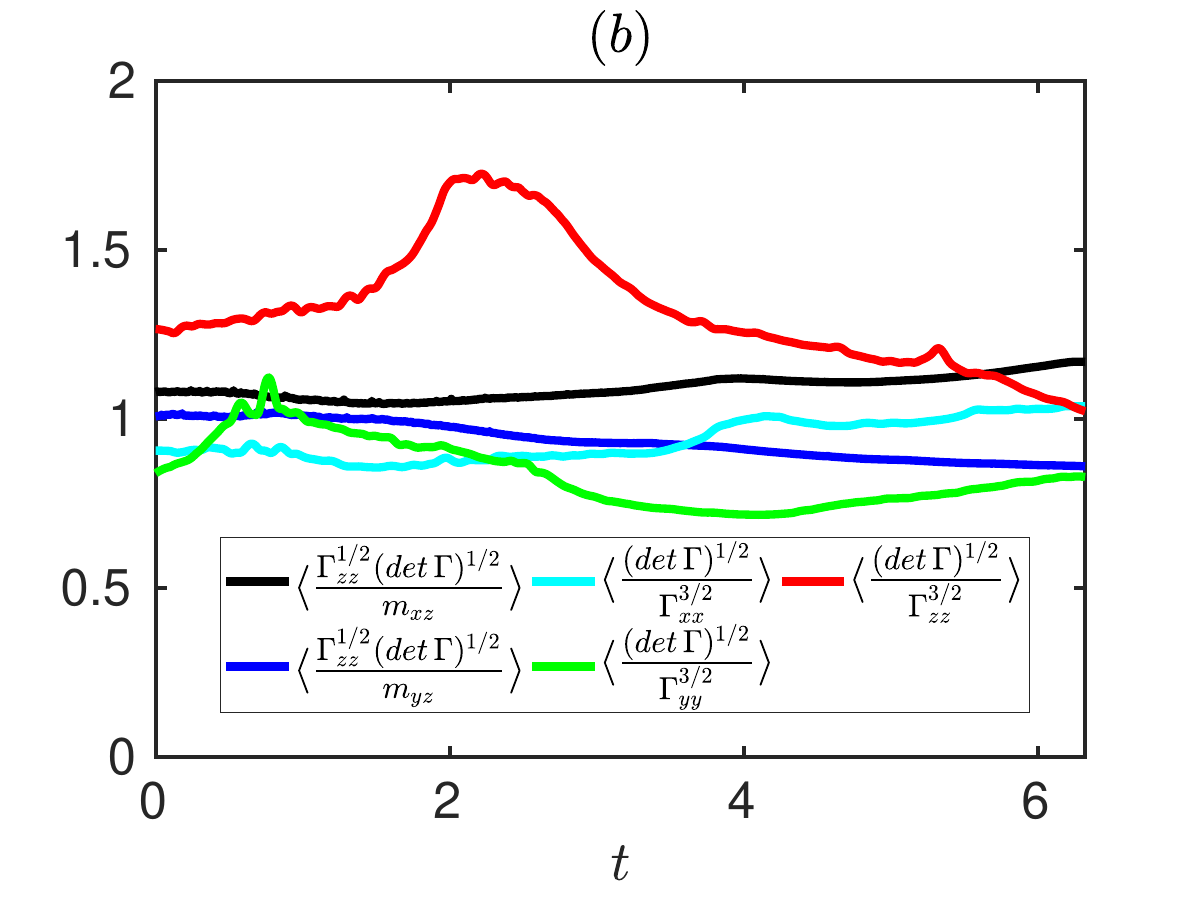}
	
	\caption{\small {\it (Color on-line)} 
	Evolution of isotropy markers~(\ref{m1})-(\ref{m3}) in the simulation of generic periodic initial flow $\mathrm{IF_{1}}$. 
	Markers~(\ref{m3}) have been smoothed using a moving average; smoothing has not been applied to other markers. 
	}
	\label{fig:fig1}
\end{figure}

\begin{figure}[t]\centering
	\includegraphics[width=9cm]{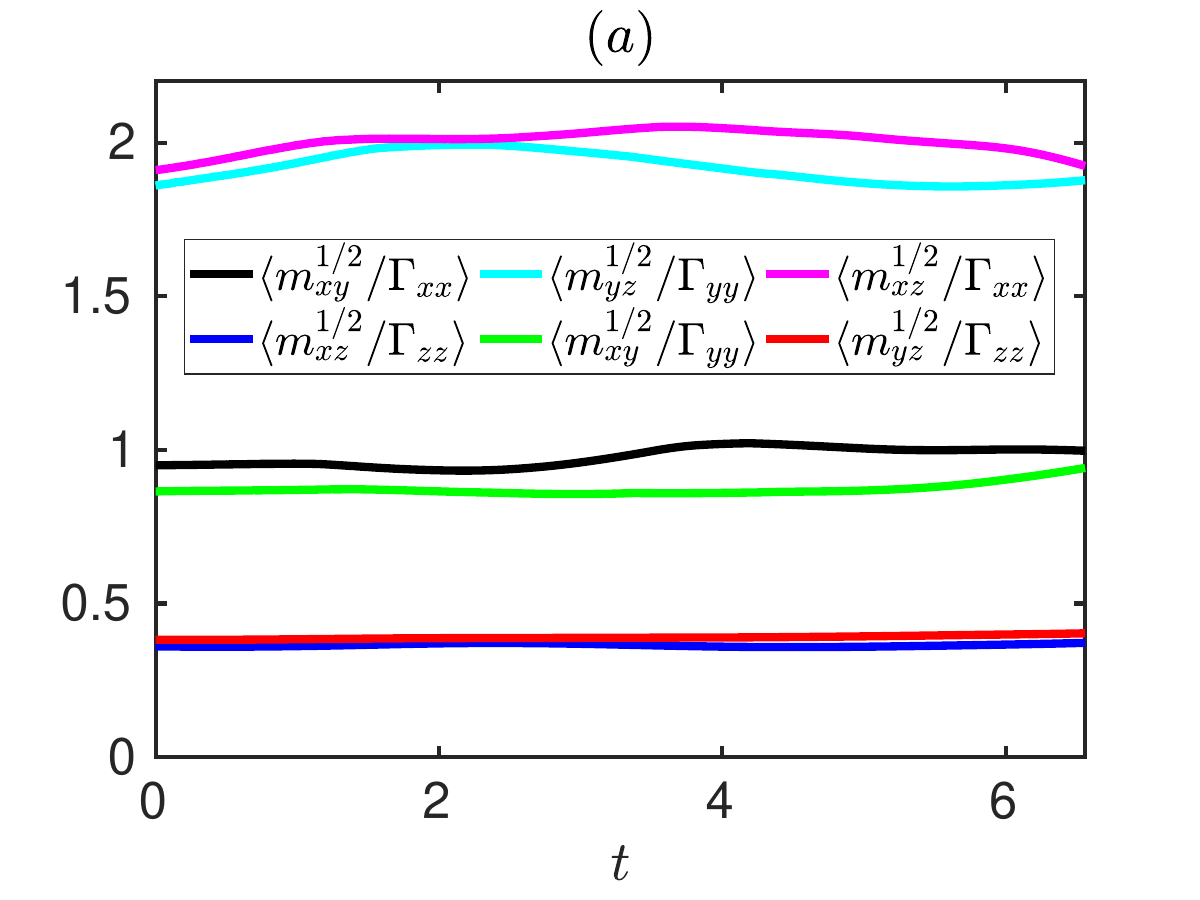}\\
	\includegraphics[width=9cm]{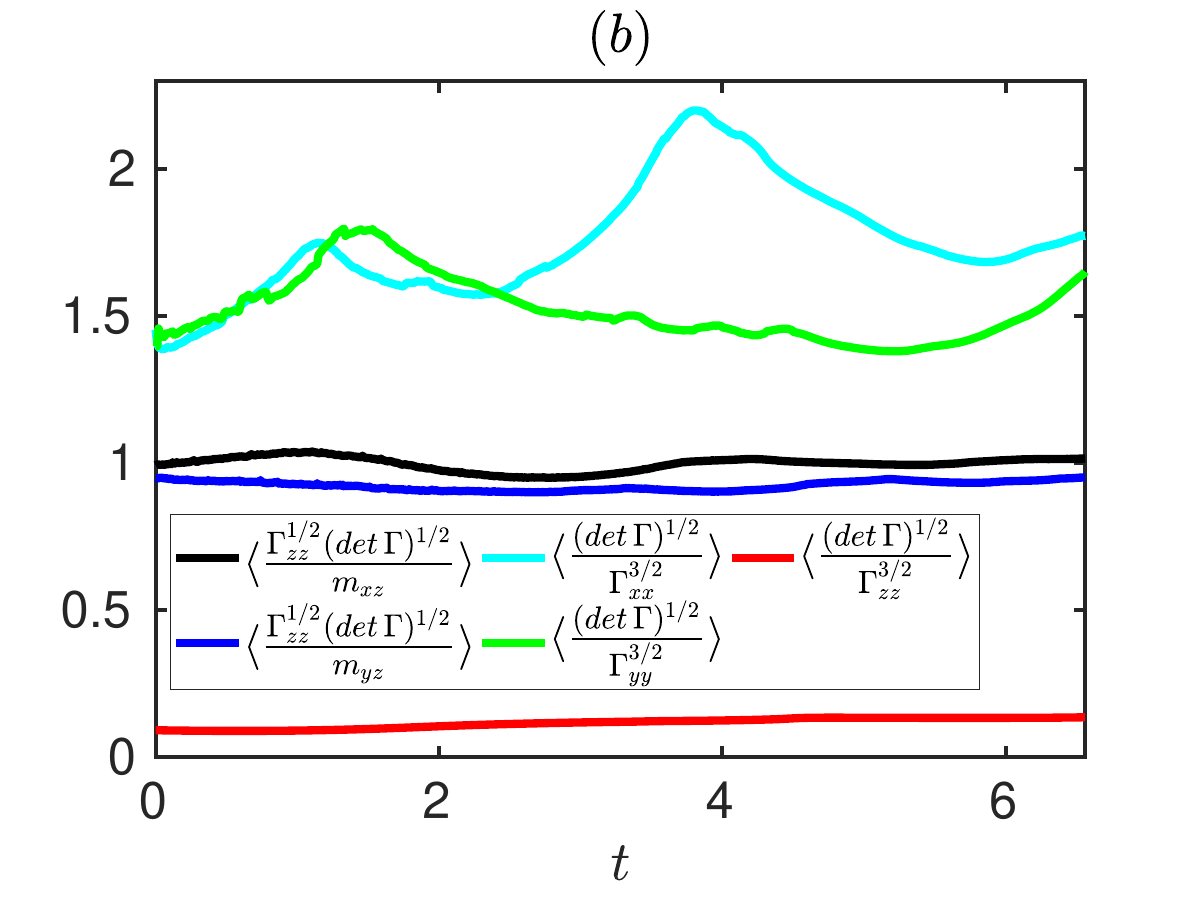}
	
	\caption{\small {\it (Color on-line)} 
    Evolution of isotropy markers~(\ref{m1})-(\ref{m3}) in the simulation of initial flow $\mathrm{IF_{0.4}}$. 
	Markers~(\ref{m3}) have been smoothed using a moving average; smoothing has not been applied to other markers. 
	}
	\label{fig:fig2}
\end{figure}

\begin{figure}[t]\centering
	\includegraphics[width=9cm]{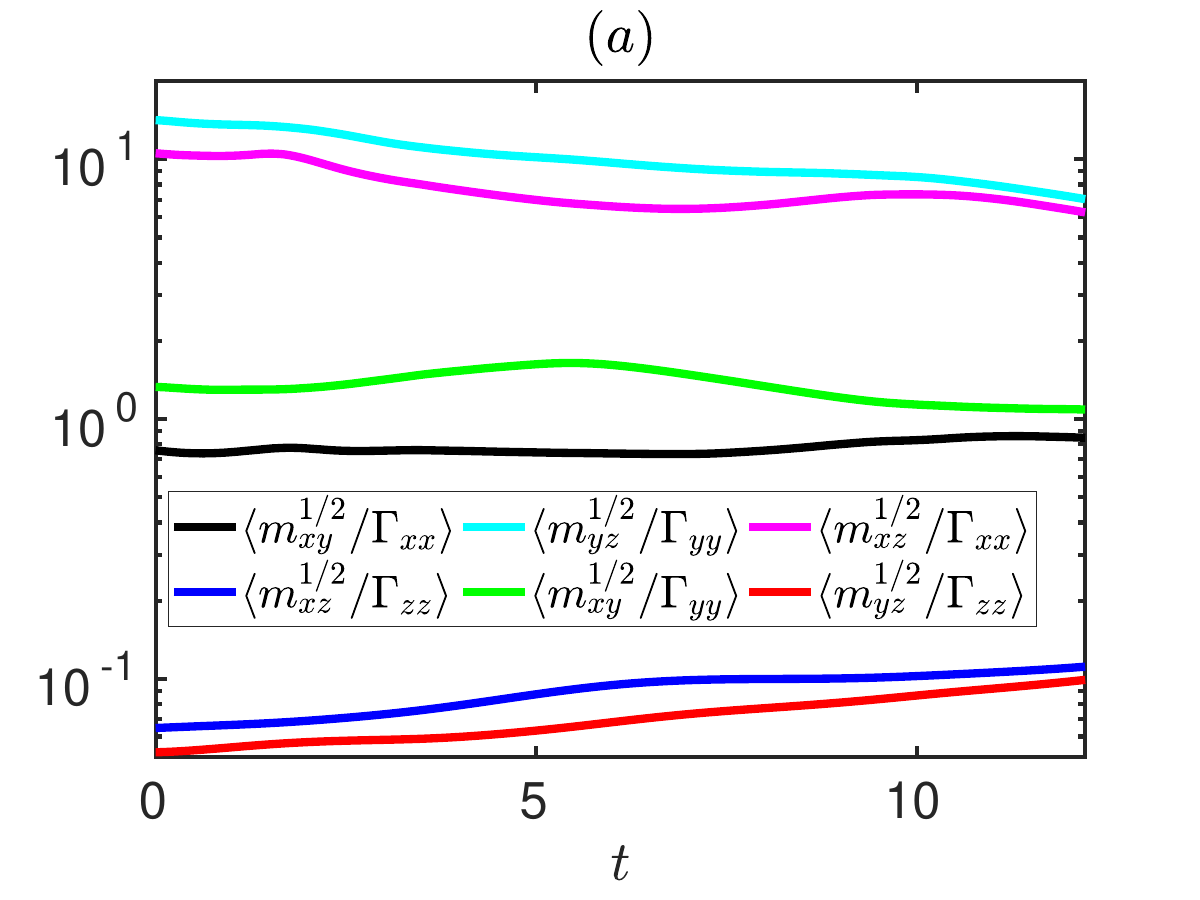}\\
	\includegraphics[width=9cm]{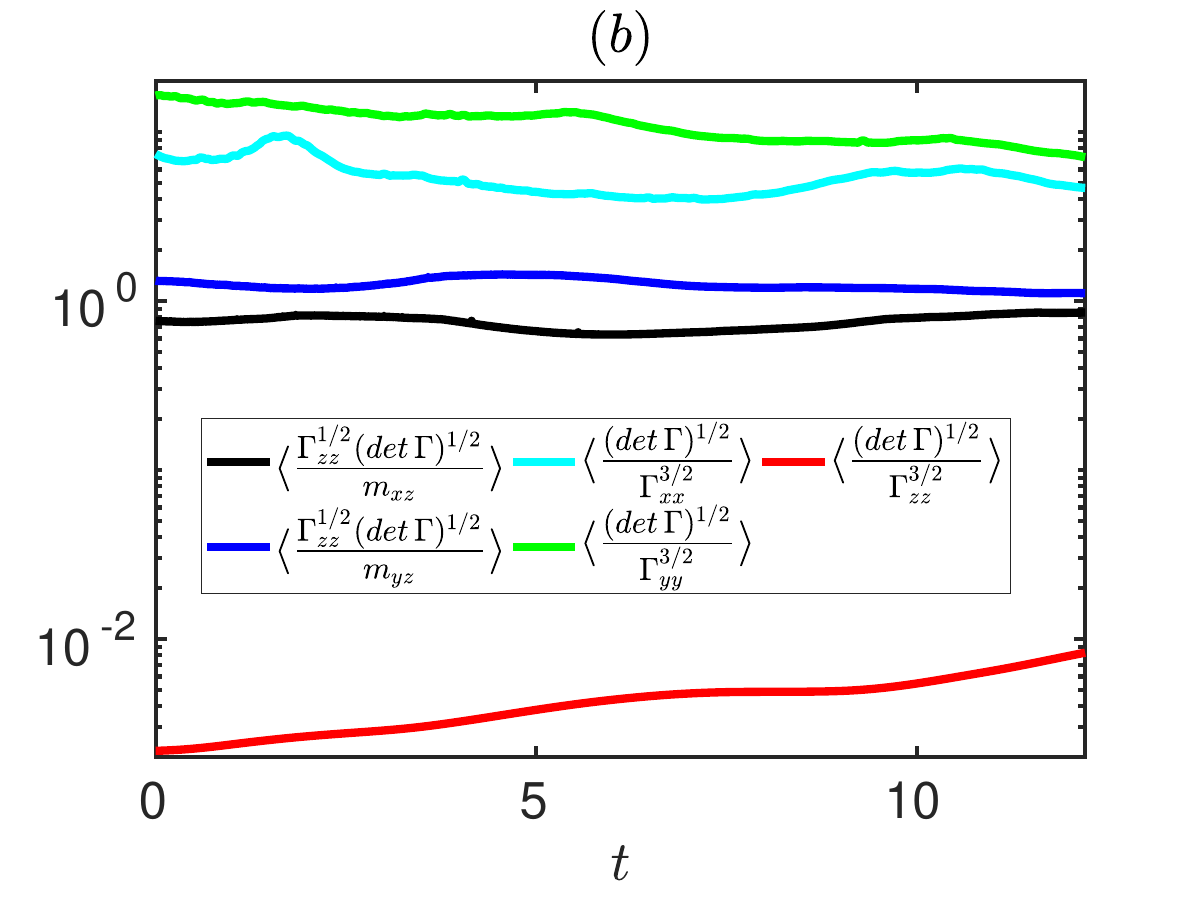}
	
	\caption{\small {\it (Color on-line)} 
	Evolution of isotropy markers~(\ref{m1})-(\ref{m3}) in the simulation of initial flow $\mathrm{IF_{0.1}}$; note the logarithmic vertical scale. 
	Markers~(\ref{m3}) have been smoothed using a moving average; smoothing has not been applied to other markers.
	}
	\label{fig:fig3}
\end{figure}

The evolution of isotropy markers~(\ref{m1})-(\ref{m3}) is shown in Fig.~\ref{fig:fig1}. 
Except for the marker $\big(\mathrm{det}\,\Gamma\big)^{1/2}/\Gamma_{zz}^{3/2}$, which briefly deviates from unity by more than 70\% and then drifts almost exactly towards unity, the other markers remain within 25\% of unity throughout the evolution time. 
Note that for markers~(\ref{m3}), indicated in panel (b) of the figure by the cyan, green and red lines, we had to apply smoothing using a moving average, since the corresponding time dependencies contained rare ``spikes'' with amplitude up to 25\% of the marker value; these spikes originate due to the denominators $\Gamma_{ii}^{3/2}$ becoming practically zero on the numerical grid near certain points. 
The smoothing has not been applied to other markers. 

We now focus on anisotropic initial flows that favor the similar orientation of the emerging pancake structures. 
Specifically, we conduct simulations for $10$ initial flows with $M=0.4$ and $a_{0}=0.7$ (group I) and another $10$ initial flows with $M=0.1$ and $a_{0}=0.9$ (group II). 
These simulations are discussed below using one representative case for each group; the corresponding initial flows are denoted as $\mathrm{IF_{0.4}}$ for group I and $\mathrm{IF_{0.1}}$ for group II. 

Figure~\ref{fig:fig2} shows evolution of isotropy markers for the simulation of initial flow $\mathrm{IF_{0.4}}$, which ended at the final time $t_{f} = 6.56$ on the grid $864\times 1024\times 1152$ and demonstrated a $9.2$-fold increase in the global vorticity maximum from its initial value $\omega_{\max}=1.1$ to its final value $10.0$. 
At the final time, we identified $31$ pancakes, and their perpendicular directions (along which the vorticity changes most rapidly) were oriented within 60$^{\circ}$ angle around the $z$-axis. 
As demonstrated in the figure, markers~(\ref{m1})-(\ref{m2}) remain practically unchanged over time, deviating during the evolution by no more than 10\% from their average values. 
For some markers, e.g., $\langle m_{xy}^{1/2}/\Gamma_{xx}\rangle$ and $\langle m_{xy}^{1/2}/\Gamma_{yy}\rangle$, these averages are close to unity, while for the others, e.g., $\langle m_{xz}^{1/2}/\Gamma_{zz}\rangle$ and $\langle m_{yz}^{1/2}/\Gamma_{zz}\rangle$, the averages differ from unity by more than two times. 
At the same time, markers~(\ref{m3}) demonstrate the greatest deviations within 10-25\% from their average values during the evolution, while the value of $\big(\mathrm{det}\,\Gamma\big)^{1/2}/\Gamma_{zz}^{3/2}$ (between $0.09$ and $0.13$) turns out to be the farthest from unity. 

The typical values of the markers can be explained by the presence of a preferred orientation for the pancakes. 
Indeed, orientation within a certain angle around the $z$-axis means that the velocity field has gradients along $z$-axis that are statistically larger~\cite{agafontsev2016asymptotic} than along the other two axes. 
Therefore, markers containing $z$-gradients in the numerator at a power greater than that in the denominator, e.g., $\langle m_{yz}^{1/2}/\Gamma_{yy}\rangle$ and $\langle m_{xz}^{1/2}/\Gamma_{xx}\rangle$, have values greater than unity. 
Conversely, if the $z$-gradients in the numerator are at a power smaller than that in the denominator, e.g., for $\langle m_{xz}^{1/2}/\Gamma_{zz}\rangle$ and $\langle m_{yz}^{1/2}/\Gamma_{zz}\rangle$, the corresponding markers have values significantly smaller than unity. 

Simulations for the other $9$ initial flows with $M=0.4$ demonstrated practically the same results, including the typical values for the markers~(\ref{m1})-(\ref{m3}). 
Only in two simulations did we observe a noticeable trend for the markers towards unity, which can be interpreted as a gradual, albeit slow, isotropization of the flow. 

Figure~\ref{fig:fig3} shows evolution of isotropy markers for the simulation of initial flow $\mathrm{IF_{0.1}}$, which ended at the final time $t_{f} = 12.21$ on a strongly anisotropic grid $648\times 648\times 2304$ and demonstrated a $6.9$-fold increase in the global vorticity maximum from its initial value $\omega_{\max}=0.97$ to its final value $6.7$. 
At the end of the simulations, we identified $49$ pancakes oriented within 20$^{\circ}$ angle around the $z$-axis. 
Due to such an orientation, it is not surprising that the typical values of the markers dependent on $z$-gradients may differ by more than $10$ times from unity. 
The marker $\big(\mathrm{det}\,\Gamma\big)^{1/2}/\Gamma_{zz}^{3/2}$ shows the maximum deviation; its value changes from $2\times 10^{-3}$ to $8\times 10^{-3}$ at the final simulation time. 
All markers that have initial values strongly deviating from unity show a noticeable trend towards unity. 
Moreover, given the logarithmic vertical scale used in the figure, this trend can be characterized as exponential-like.

We observed the same results for simulations that started from the other $9$ initial flows with $M=0.1$. 
Moreover, we repeated simulations for another two initial flows studied extensively in~\cite{agafontsev2015development,agafontsev2016asymptotic,agafontsev2018development}, and again came to the same conclusions. 


\section{Conclusions}
\label{Sec:Conclusions}

We have systematically examined the evolution of isotropy markers for periodic flows that initially represent a mixture of a shear and random flows. 
The shear flow is introduced to influence the orientation of the emerging pancakes: from fully random, when this flow is absent, to almost unidirectional, when it dominates. 
We have found that, for a random initial flow, the isotropy markers remain sufficiently close to unity at all times. 
When the shear flow is moderately present, the isotropy markers practically do not change with time and some of them have values deviating from unity by a few times. 
When the shear flow dominates, all of the initial flows we have studied led to a nearly unidirectional pancake orientation, while the isotropy markers, initially deviating from unity by orders of magnitude, demonstrated a slow drift, similar to an exponential decay, towards unity. 

These observations allow us to draw several conclusions. 
First, in all the cases considered, the isotropy markers exhibit a relatively slow evolution over time, indicating that the flow does not have a tendency towards spontaneous symmetry breaking. 
Secondly, strongly anisotropic initial flows undergo a process of slow isotropization with time. 
Importantly, this process starts without the viscous scales getting exited and continues despite the similar orientation of the emerging pancake structures. 
What flow behavior causes this phenomenon remains to be determined. 
For instance, isotropization may occur due to the appearance of new pancakes and the gradual widening of the angle within which they are oriented. 
However, our numerical resources are insufficient to test this hypothesis reliably. 
It may also occur due to the thinning and slight warping of the older pancakes~\cite{agafontsev2016asymptotic}. 
Note that the pancakes remain stable~\cite{agafontsev2021stability} and are not destroyed until the viscous scales come into play~\cite{krstulovic2024initial}. 
We also suggest that isotropization takes place for moderately anisotropic flows as well; however, its observation requires much longer evolution times, i.e., much greater numerical resources.

\corr{In the present paper, we have studied evolution of isotropy markers constructed from the velocity gradients tensor. 
As a possible direction for future research, one may consider isotropy markers built from other second-rank tensors, e.g., for which the influence of the small-scale motion is significantly stronger. 
Another possible direction is to examine the influence of flow helicity on the evolution of pancake structures and isotropy markers. 
According to the exact pancake solution~\cite{agafontsev2016asymptotic}, the helicity of a pancake equals zero. 
However, the background flow and the transient flow between the background and the pancakes may have non-zero helicity. 
This may influence the emergence and evolution of pancakes and, consequently, the behavior of isotropy markers. 
However, studying this question requires a large number of simulations for different levels of helicity of the initial flows, which significantly exceeds our current computational resources.}
\\

{\it Acknowledgments.} 

The authors are grateful to E.A. Kuznetsov, V.A. Sirota and K.P. Zybin for valuable discussions. 
Simulations have been performed at the Novosibirsk Supercomputer Center (NSU). 


%

\end{document}